# High-speed x-ray imaging with single-pixel detector

**Authors:** O. Sefi[1], Y. Klein[1], E. Strizhevsky[1], I. P. Dolbnya[2], and S. Shwartz[1]

[1] Physics Department and Institute of Nanotechnology, Bar-Ilan University, Ramat Gan, 52900 Israel

[2] Diamond Light Source Ltd, Harwell Science & Innovation Campus, Didcot, Oxfordshire OX11 0DE, UK

## Abstract

We demonstrate experimentally the ability to use a single pixel detector for two-dimensional high-speed and high-resolution x-ray imaging. We image the rotation of a spinning chopper at 100-kHz at spatial resolution of about 15 microns by using the computational ghost imaging approach. The technique we develop can be used for the imaging of high-speed periodic dynamics or periodically stimulated effects with a large field of view and at low dose.

# Introduction

X-ray imaging methods are used for numerous applications in a variety of areas ranging from basic science to medicine and security. The main advantage of x-rays for imaging is their unique capability to penetrate through surfaces, which are opaque to other commonly used wavelengths, such as visible and infrared. While most x-ray pixelated detectors are used for static applications, there is a growing need for detectors with combined high spatial resolution and at high frame rates. These detectors will open a new horizon for the application of x-rays for measurements of dynamics of systems, such as acoustic waves in matter, dynamics of phase transitions, and medical imaging. This innovative function will be essential for the understanding of numerous phenomena and processes and can lead to novel applications.

However, similar to other regimes of the electromagnetic spectrum, two-dimensional x-ray detectors are much slower than single pixel detectors, while single pixel detectors cannot provide the spatial resolution directly. Even with the improvements in detector technologies, the fundamental tradeoff between the number of pixels and the readout time is a major setback that detains the remedy of this challenge. For example, for detectors with 1360 x 1080 pixels the frame rates are about 10 frames per second [1]. Other detectors with fewer pixels can be faster, but they exhibit either poor resolution or small field of views [2,3]. We note that modern detectors can use fast electronic shutters for short acquisition times on the order of only a few microseconds, but their frame rates, field of views and spatial resolution are still quite limited and they are very expensive [4]. While other x-ray cameras exhibit frame rates in the range of MHz, the number of recorded frames is limited [5]. Sophisticated imaging methods, such as pulse isolation and pump-probe imaging [6–9], have shown some success as well. However, those methods require very specific time structures of the sequence of synchrotron radiation pulses. Additionally, none of those methods can be performed with x-ray tubes, which are the most abundant x-ray sources.

A possible route for a solution that offers high frame rate imaging at high resolution can be based on the method of ghost imaging (GI). This is a lens less imaging technique, in which intensity fluctuations are intentionally added into the input beam (usually by a diffuser). In the traditional scheme, after the diffuser, the beam is split into two portions with identical illumination patterns, by a beam-splitter [10–13]. A portion of the beam (usually about 50 % of the intensity), which is called the reference beam, propagates freely and is collected by a two-dimensional detector that measure the spatial distribution of the intensity fluctuations. The second portion of the beam, which is called the test beam, impinges on the object and is then collected by a single pixel detector, which is mounted behind the object. The image is reconstructed by correlating the reference and the test beams for many different realizations of the illumination patterns, which are obtained by raster scanning the diffuser. In another approach for the implementation of GI [14–19], the measurements are performed in two steps: first, the distributions of the intensity fluctuations for the various realizations are measured and recorded by the two-dimensional detector in the absence of the object. In the second step the object is inserted, the two-dimensional detector is replaced by the single pixel detector, and the measurements of the test beam are recorded for the same positions of the diffuser. After the two sets of measurements are completed, the image is reconstructed by correlating the measurements from each realization. The method has been demonstrated mainly with optical radiation and recently with x-rays [11–13,15–17,20–23], atoms [24], electrons [25], and neutrons [26]. In addition, far field GI [17,27,28], GI in the time domain [29], ghost polarimetry [30], ghost tomography [13], and ghost spectroscopy [31] have been demonstrated as well.

While in the standard GI the object is static, the method can be extended for the imaging of motions of objects that vary periodically by using the following procedure similar to the procedure shown by Zhao et al. with optical radiation [32]: The measurement of the reference beam is done by a slow pixilated detector with high spatial resolution. In the second step, we measure the test beam. This is done by measuring the intensity at the single-pixel detector for the entire period of the motion of the object at the high sampling rate allowed by the single-pixel detector for one position of the diffuser. Since this system measures only the signal of a single-pixel detector, the frequency of the recorded data can exceed tens of MHz. Synchronization between the period of the motion of the object and the data acquisition system is essential. We repeat this measurement for the same set of realizations as was done with the reference beam by scanning the diffuser. Next, we reorder the recorded data sets to create a new sequence of data sets where each includes the measurements of the reference and the test beams for all of the positions of the diffuser for one frame (one position of the object). We do that by using the information from the synchronization signal. We then corelate the reference and the test beams for each frame separately to reconstruct the image as is dome with the static GI. In the final step, we construct the sequence of the frames by concatenating the individual frames.

In this work we demonstrate experimentally, for the first time, the possibility to use x-ray GI to image fast dynamics. We use the method to image the motion of a chopper spinning at 200 Hz where the frame rate is 100 kHz and at spatial resolution of about 15 microns and with a field of view of 0.6 mm × 0.6 mm. The procedure we demonstrate can be used for the study of dynamics for a variety of phenomena with x-rays.

## Experiment & procedure

We performed the experiment on the B16 beamline at the Diamond Light Source. The schematic description of the experimental setup is shown in Fig. 1 (a)-(b). We exploit the two-step approach for the implementation of GI as we described above. We use a monochromatic beam at 9 keV on which we place a squared aperture with the size of 0.6 mm × 0.6 mm. The beam propagates through a sandpaper with an average feature size of about 10 μm, that we use as a diffuser, mounted on two-dimensional stages. We measure and record the intensity fluctuations with a Photonic Science X-ray MiniFDI camera for which the pixel size is 6.5 μm and the frame rate is 7.5 frames per second. An explanatory image of the reference beam is shown in Fig. 1 (c). After we measure the diffuser, we insert the object, a ThorLabs optical chopper rotating at chopping frequency of 200 Hz with jitter of 0.2 deg per cycle and the velocity of the blades of the chopper is about 0.6 m/sec. We use the chopper output signal for synchronization. For the measurements of the test beam we use an avalanche photodiode detector (APD). We record the data with a NI-DAQ USB X SERIES data acquisition card at the sampling rate of 100 kHz. An example for the test and for the chopper signals can be seen in Fig. 1 (d).

For the image reconstruction we use the compressive sensing approach and use the TVAL3[1] reconstruction algorithm [18,33]. In this approach, a prior information on the structure of the image is used. Since many objects in nature have a basis in which their image is sparse, meaning that the image has many components, which are zero or close to it, one can reconstruct the image of such an object by finding the image $T_{CS}(x,y)$, which minimizes the $L_1$-norm in the sparse basis, namely:

$$T_{CS} = \min_{T'} \|\Psi\{T'(x,y)\}\|_{L_1}$$

$$subject\ to\ \int dxdy I_r(x,y)T'(x,y) = B_r\ \forall_{r=1...N}$$

here, $\Psi$ is the transform operator to the sparse basis, $I_r$ is the r[th] illumination pattern and total number of realizations is $N$. $B_r$ is the r[th] single-pixel test measurement. The meaning of this condition is that the algorithm favors an image, which minimizes the sum of absolute values of its components in the sparse basis. This method is very useful in reducing the required number of realizations.

---

[1] We used the isotropic TV/L2+ model with the parameters $\mu = 2^8$ and $\beta = 2^6$.

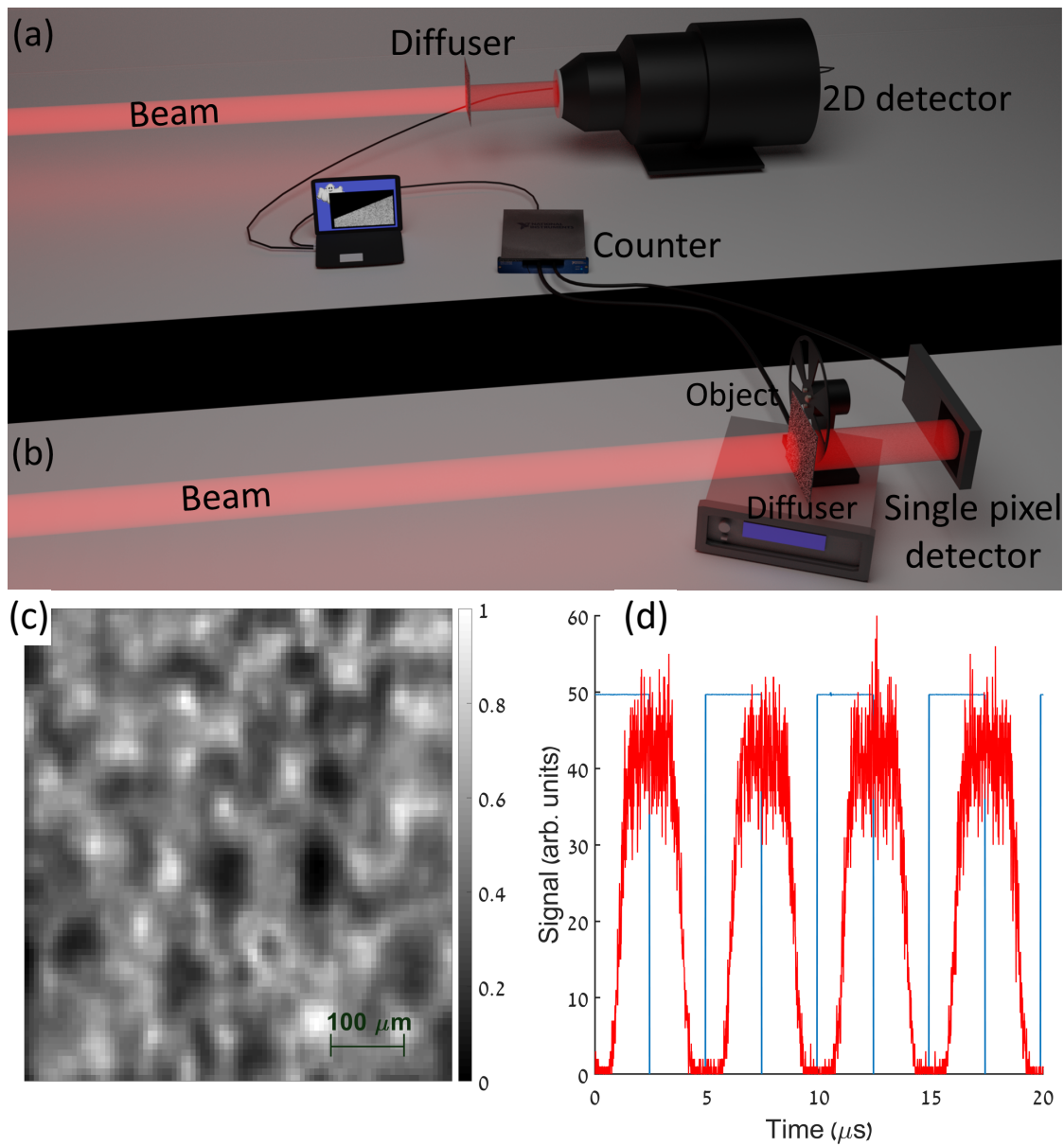

Figure 1 – The experimental setup. (a) The setup for the reference beam measurements: The intensity fluctuations are introduced into the beam by a motorized diffuser. Afterwards, the intensity distribution is measured by a 2D x-ray detector. (b) The setup for the test beam measurements: The intensity fluctuations are introduced into the beam by the diffuser and the beam impinges on the object. The beam that is transmitted through the object is

### Results & discussion

We start by showing several snapshots from the reconstructed movie of the optical chopper, which are shown in Fig. 2. The number of pixels in each frame is 8550 and the frame rate is $10^5$ frame per second. In each test realization we sum over 40 periods of the chopper to enhance the test signal and improve the signal-to-noise ratio (SNR). We use 4900 realizations per each frame with an average of $9.5 \cdot 10^4$ counts per realization. In the upper left frame, the blade of the chopper blocks almost the entire beam except from a small area near the bottom right corner. The next panel shows the 6th later frame where the chopper blocks a smaller portion of the beam. The rest of the frames show the motion of the chopper at measurement times that are indicated in Fig. 2 until the bottom right panel where the blade blocks the beam except from a small area near the upper left corner. The corresponding movie is provided in 'HighDose_100kHz.avi'.

Note that the edge of the blade shown in the images is smeared over about 10 pixels, which correspond to 65 microns. There are several reasons for that blurring: 1. Since we average over multicycles, the jitter of the chopper, which is about 0.2 deg per one blade cycle, is responsible for smearing of 10 microns. 2. The resolution of the x-ray two-dimensional detector that we use for the reference beam leads to additional blurring of about 20 microns. 3. We believe that the reconstruction procedure errors are the reason for the additional smearing of about 30 microns and this can be improved by improving the system. The blades move about 6 microns (about one pixel) in 10 µs.

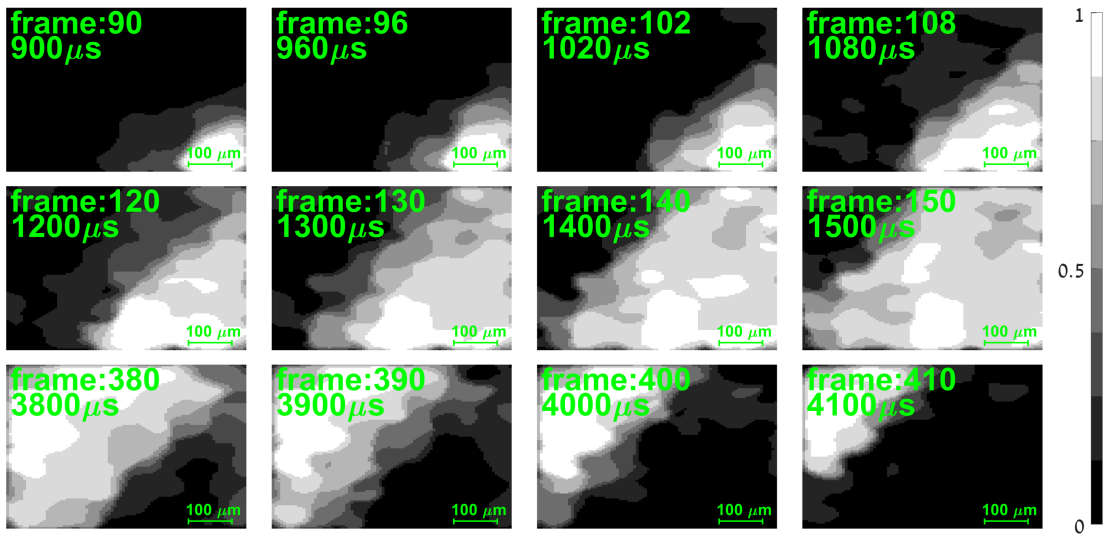

*Figure 2 – Reconstructed movie frames of an optical chopper rotating at 200 Hz. The movie is recorded at a frame rate of $10^5$ frames per second. The average flux in each test measurement is $\sim 9.5 \cdot 10^4$ counts. Each frame has $\sim 8 \cdot 10^3$ pixels with pixel size of 6.5 microns. The intensities of the images are normalized, see details in the text.*

After demonstrating the feasibility of high-speed x-ray imaging, we explore the quality of the reconstruction and its dependence on the number of realizations. We show the dependence of the SNR on the number of realizations in Fig. 3. The blue dots are the SNR that is calculated from the measured data and the red solid line is the fitting function $f = a\sqrt{N}$ with $a = 0.1033$. Here, we use the standard definition of $SNR = \mu_I/\sigma_I$, where $I$ is a vector of the intensities of several pixels in the illuminated area of the image, $\mu_I$ is the ensemble average, and $\sigma_I$ is the standard deviation. The vertical error bars indicate the counting statistics, which are estimated by the standard error of the SNR, namely $\left(1/I + \mu^2/\sigma^3\sqrt{I}\right)^{1/2}$. Theoretically, the SNR of standard GI scales as the squared root of the number of realizations divided by the number of the used pixels, namely: $SNR \propto \sqrt{N/p}$, where p is the number of pixels [22]. Since we use the compressive sensing approach the SNR is about ten times better [18].

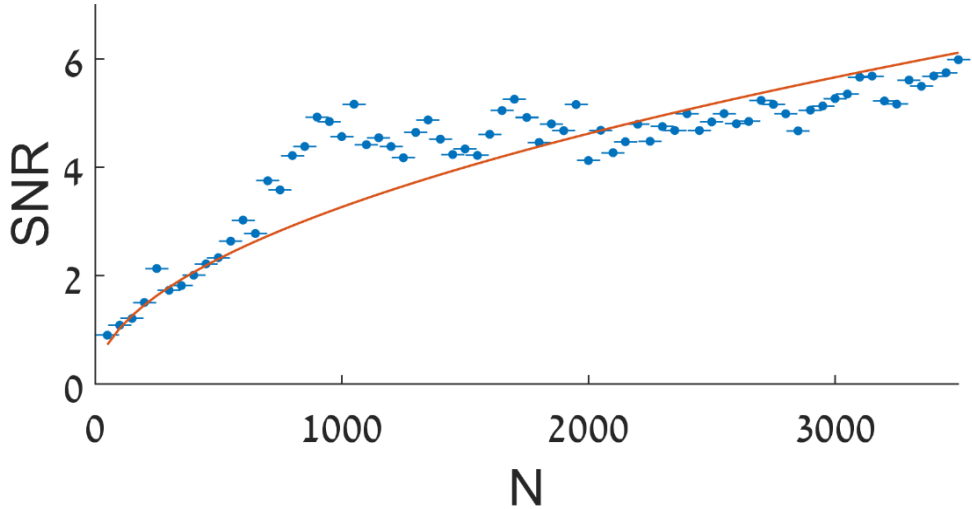

*Figure 3 – The SNR as a function of the number of realizations N for frame 112 of the movie. The blue dots are experimental results and the red solid line is the analytic fitting function $a\sqrt{N}$. The error bars indicate the counting statistics. See detail in the text.*

## Low dose imaging

One of the main potential advantages of GI is the opportunity for low dose imaging that the technique offers [12,16,22,34]. We explore this ability with high-speed imaging by investigating the dependence of the SNR on the number of the photons in the test beam. The results are summarized in Fig. 4(a) where the blue dots indicate the SNR calculated from the measured data and the error bars indicate the counting statistics. The SNR and the vertical error bars are calculated as in Fig. 3. The insets (b)-(d) are the corresponding reconstructed images for average flux of 30, 375 and 1160 counts respectively.

The dependence of the SNR on the number of photons presented in Fig. 4 can be separated into two distinct regions. In the low flux region, up to about 1000 counts, the SNR scales as $a\sqrt{N}$ with $a = 0.1033$, like the dependence shown in Fig. 3. However, at higher flux the dependence on the number of photons is weaker. In this region the coefficient $a$ is reduced by about factor of ten. This dependence indicates that in the high flux regime the statistical error is negligible and other sources for noise, that probably can be reduced, dominant the noise in this regime. In the low dose region however the main source of noise is shot noise, which is standard quantum limit and it is the lowest achievable noise with classical sources.

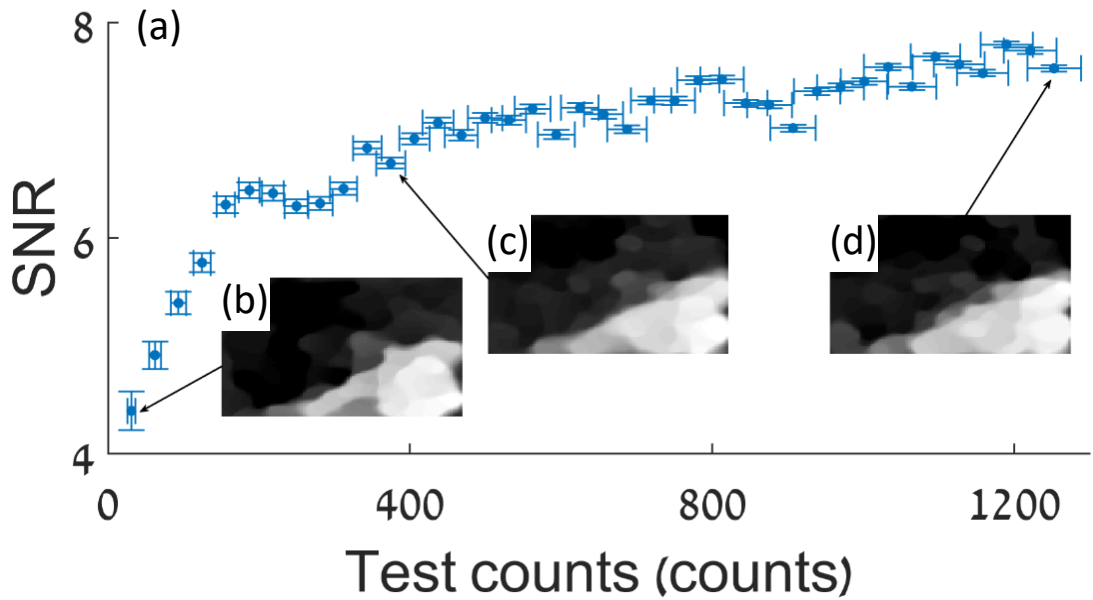

*Figure 4 - (a) Dependence of the SNR on the flux of the test beam corresponding to frame 112 in the movie. The horizontal error bars indicate the counting statistics. In the insets (b-d) we show the reconstructed images, which correspond to flux of 30, 370, and 1150 counts respectively.*

We further illustrate the low dose imaging capabilities by showing snapshots from the reconstructed movie under low dose conditions in Fig. 5. Here, we use a single cycle of the motion of the chopper per test measurement. Consequently, the average number of photons per measurement is 24 counts. We use 4900 realizations and reconstruct images, which include 8500, pixels and measure on average about 15 counts per frame per pixel. The frames that we show here are the same as in Fig. 2 and they describe the same dynamics. This result shows the feasibility to retrieve reasonable results even with extremely small amount of radiation, which is orders of magnitude smaller than the radiation levels used in traditional direct imaging. This capability is extremely important for high-speed imaging, where the exposure times are inherently small thus inherently associated with a small number of photons per frame. The corresponding movie is provided in 'LowDose_100kHz.avi'.

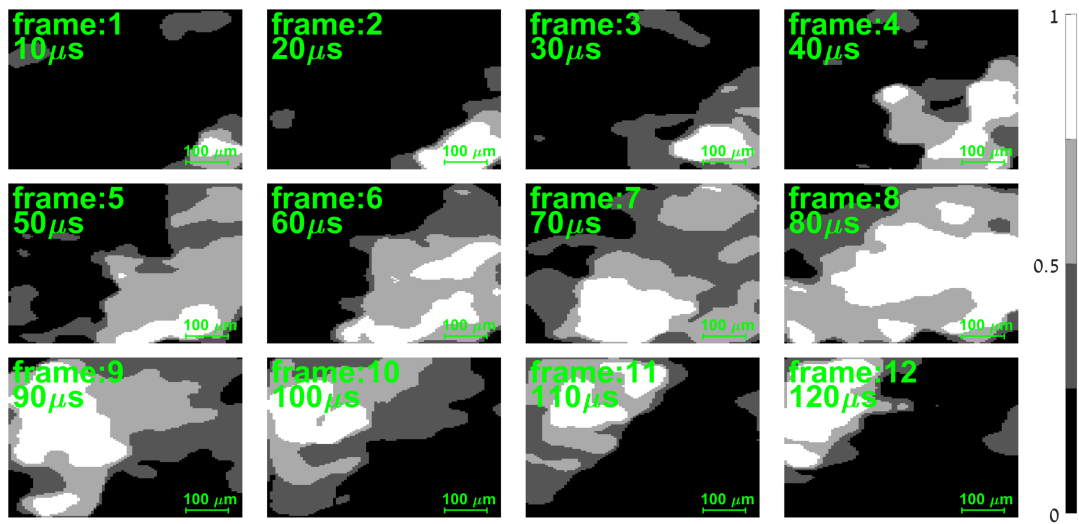

*Figure 5 – Low-dose reconstructed movie frames of an optical chopper rotating at 200 Hz. The average number of photons in each test measurement is 24. Each frame includes $\sim 8 \cdot 10^3$ pixels with pixel size of 6.5 microns. The intensities of the images are normalized, see details in the text.*

Before concluding, we discuss two important aspects related to the spatial and temporal resolution of the system. Of importance, the spatial resolution is determined by the resolution of the measurements of the reference beam. Here, we used an x-ray pixilated detector with a limited resolution to perform those measurements. It is possible to improve the resolution by using a free-propagation calculation from a diffuser with known smaller features that can fabricated with Nanotechnology techniques, as is done in computational GI [14,15]. As for the temporal resolution, in our system the frame rate was limited by the DAQ sampling frequency. However faster DAQs are available, with sampling frequencies up to 100 MHz. The next factor which limits the frame rate is the detection time of the APD, which is about 10 ns for standard APDs.

## Conclusions

We have demonstrated the ability to use the method of GI for high-speed large field of view x-ray imaging at high-resolution. We have also shown that the method can be used for low dose high speed imaging. The procedure and results we have described opens new possibilities for the study of fast dynamics of processes, which can be triggered periodically. The method is not limited to transmission measurements and can be used for high-speed phase sensitive measurements [23,35] and to incoherent diffraction imaging [17,23]. Finally, we note that the high-speed GI method can be implemented with conventional x-ray tubes, hence can be useful for medical imaging and industrial applications.

## Funding

The research leading to this result has been supported by the project CALIPSOplus under Grant Agreement 730872 from the EU Framework Programme for Research and Innovation HORIZON 2020. Y.K. gratefully acknowledges the support of the Ministry of Science &Technologies, Israel.

## Acknowledgments

We thank Diamond Light Source for access to Beamline B16 (Proposal No. MM23604), which contributed to the results presented here.